\documentclass[12pt]{article}

\usepackage{graphics}

\newcommand{\ket}[1]{\vert#1\rangle}

\newcommand{\ceil}[1]{\lceil{#1}\rceil}

\title{Quantum Arithmetic on Galois Fields}

\author{St\'{e}phane Beauregard, Gilles Brassard, Jos\'{e} Manuel Fernandez}

\date{11 April 2002}

\begin{document}

\maketitle

\begin{abstract}
In this paper we discuss the problem of performing elementary finite field arithmetic on a quantum computer.  Of particular interest, is the controlled-multiplication operation, which is the only group-specific operation in Shor's algorithms for factoring and solving the Discrete Log Problem.  We describe how to build quantum circuits for performing this operation on the generic Galois fields GF($p^k$), as well as the boundary cases GF($p$) and GF($2^k$).  We give the detailed size, width and depth complexity of such circuits, which ultimately will allow us to obtain detailed upper bounds on the amount of quantum resources needed to solve instances of the DLP on such fields.
\end{abstract}

\section{Introduction}


The most significant event in the short history of Quantum Computing is the discovery of an efficient algorithm for factoring integers by Peter Shor in 1994.  The algorithm was initially described at a high level, and it assumed the existence of an efficient quantum black box capable of computing integer modular exponentiation, i.e. computing $a^x \bmod N$, given an integer $x$, and previously known (``hardwired'') integers $a$ and $N$, the latter being the integer that we want to factor.  Shor did not bother to describe in detail such a black box, as it is trivial to show that such a box exists.  There are several classical circuits that compute them efficiently, and any of those could in principle be transformed into a quantum circuit, also of polynomial size.


While it is clear that the overhead of such a conversion is always represented by a bounded degree polynomial, the question arises of exactly how small it can be.  Given the fact that building large scale quantum computers, even of a few hundred qubits, represents a formidable technological challenge, it becomes paramount to know exactly how many qubits, and also exactly how many operations are required to construct such black boxes.  The importance of this knowledge cannot be overemphasized, given the potential cryptanalytic applications that an efficient factoring algorithm can have.

The exact complexity of performing modular exponentiations will depend on the complexity of performing simpler arithmetic operations such as addition and multiplication.  In fact, in the context of Shor's algorithm, the black box for exponentiation can be substituted for by a limited number of black boxes performing controlled modular multiplications, where one of the two factors is a previously known (i.e. ``hardwired'') value $a$.  Thus, once we have established the exact complexity of implementing these controlled multiplications, it becomes in turn possible to determine the exact complexity of the overall quantum factoring algorithm.

In the case of factoring, we are dealing with integer arithmetic and the topic of quantum integer and modular arithmetic has already been well studied, and satisfactorily resolved.  


Nonetheless, in order to fully consider the cryptanalytic potential of quantum computers, one must also consider the Discrete Log algorithm introduced by Shor at the same time as his factoring algorithm.  This algorithm was also described at a high level in terms of a quantum black box computing double exponentiations, i.e. obtaining values $a^x b^y$, given integers $x$ and $y$ and for fixed, multiplicative group elements $a$ and $b$.  As is the case for simple exponentiation, one can show that such ``efficient'' quantum circuits exist for implementing these other kind of black box, but the same questions arise about exactly how many resources are needed to build and evaluate them.  Furthermore, it can also be shown that these double exponentiation can be substituted with a known, exact number of controlled multiplications.

Nonetheless, the situation is made more complex by the fact that the Discrete Logarithm Problem (DLP) can be defined on any commutative group.  Thus, in addition to consider integer modular arithmetic, one must also consider quantum arithmetic on suitable representations of these other groups.  Of particular interest is the DLP defined on the multiplicative groups of the Galois Fields, again given its obvious cryptanalytic applications.  Also in that category are groups based on elliptic and hyper-elliptic curves.  While arithmetic of points on an elliptic curve is quite different from arithmetic on elements of Galois Fields, it is common to define these curves on a vector space over the Galois Fields.  In either case, the importance of knowing the exact complexity of performing Galois Field arithmetic, and in particular that of performing controlled multiplications is made obvious.  

In this paper, we study precisely this question.  In the next sections, we will study independently the three different cases of Galois Fields, i.e. GF($p$), GF($2^k$), and GF($p^k$), where $k$ is an integer, and $p$ a prime bigger than 2.

\section{Quantum Arithmetic in GF($p$)}

We first consider the arithmetic in the Galois field GF($p$), where $p$ is an $n$-bit prime integer. The Galois field GF($p$) is isomorphic to the integers modulo $p$. We will devise a circuit to implement the controlled multiplication modulo $p$ on a quantum computer. It is important to realize at this point that the value $p$ is a classical value and can be hardwired in the circuit. Moreover, the value $a$ by which we will multiply can also be hardwired in the circuit since it will be given as a classical input.

The most straightforward way to get a multiplication circuit is by successive addition of classical values modulo $p$. This method also seems to be the most qubit-efficient one. To get to this successive addition circuit, we begin with the simple addition of a classical value to a quantum register.

\subsection{The adder gate for integers}
\begin{figure}[htb]
\begin{center}
\includegraphics{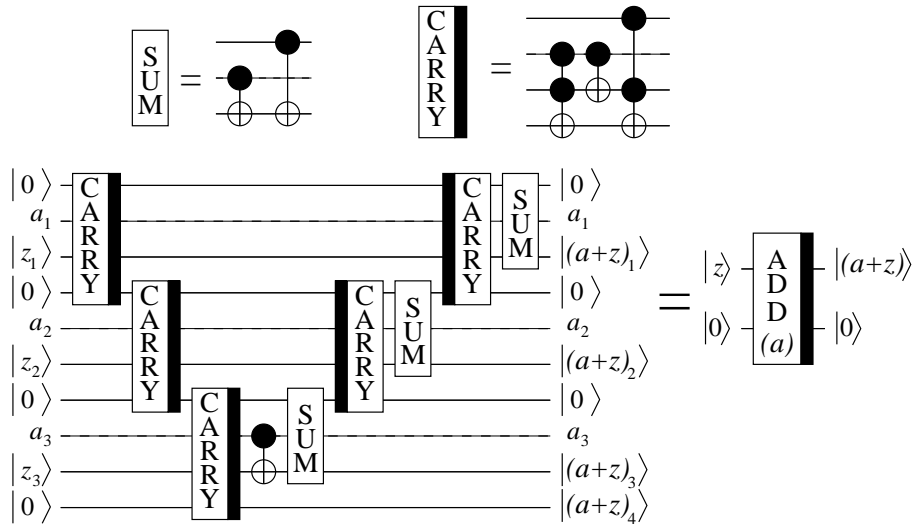}
\caption{\small{The carry-sum adder of Vedral, Barenco and Ekert modified to add a classical value ($a$) to the quantum register $\ket{z}$}}\label{carrysum}
\end{center}
\end{figure}

The adder gate for integers is simply a circuit that adds a classical value to a quantum register. We consider two ways of doing this. The first is an adaptation of the carry-sum adder of Vedral, Barenco and Ekert~\cite{VBE} (Figure~\ref{carrysum}).

This circuit requires $2n$ qubits to add a value without overflows because the first qubit of figure~\ref{carrysum} is not needed. It uses O($n$) elementary gates in linear depth.

\begin{figure}[htb]
\begin{center}
\includegraphics{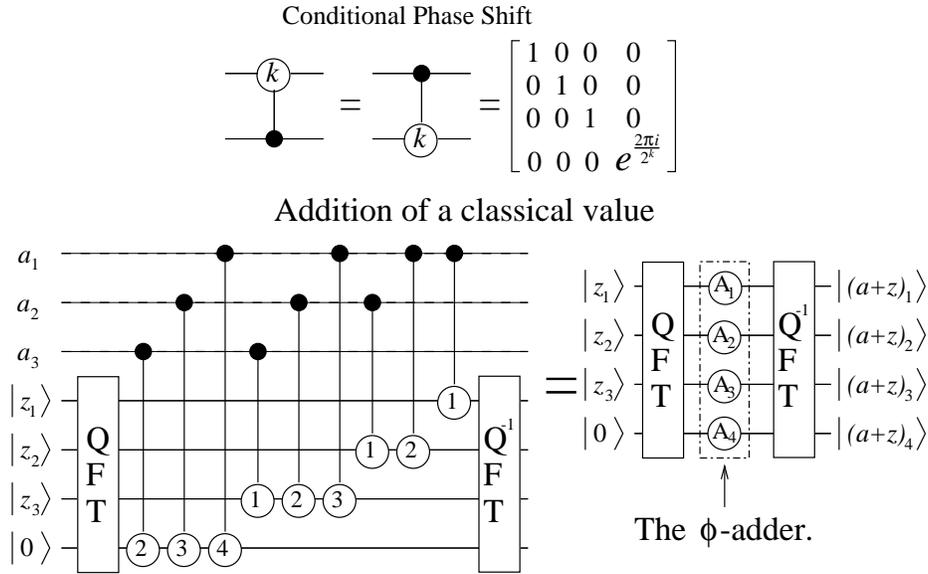}
\caption{\small{The $\phi$-adder. The fact that the value $a$ is classical helps to simplify the adder.}}\label{phiadd}
\end{center}
\end{figure}

The second method uses an adder from Draper~\cite{Draper} that we will call the $\phi$-adder(Figure~\ref{phiadd}). The $\phi$-adder takes the quantum Fourier transform of a qubit register $\ket{z}$ to the quantum Fourier transform of the sum $z+a$, where $a$ is a classical value hardwired in the $\phi$-adder. The advantage of this method is that it does not need extra qubits for carries. Furthermore, the fact that we only need to add a classical value helps to simplify the $\phi$-adder. However, a quantum Fourier transform has to be applied to the quantum register before and after the $\phi$-adder, so we end up using more elementary gates than with the carry-sum adder.

This circuit requires only $n+1$ qubits to add a value without overflows as in figure~\ref{phiadd}. What we call the $\phi$-adder does not include the QFTs since they are often not needed. The $\phi$-adder requires O($n$) elementary gates in constant depth if we exclude the QFTs. If the $\phi$-adder has to be controlled be another qubit, its depth becomes linear.

We need O($n^2$) elementary gates in depth O($n$) to implement the exact QFT on an ($n+1$)-qubit register.

\subsection{The adder gate for GF($p$)}

\begin{figure}[htb]
\begin{center}
\includegraphics{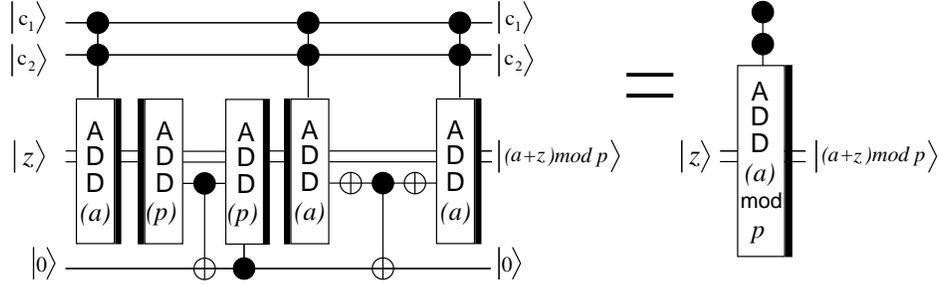}
\caption{\small{The adder for GF($p$) controlled by qubits $c_1$ and $c_2$. The value $a$ and the input $\ket{z}$ are both smaller than $p$.}}\label{ccaddmod}
\end{center}
\end{figure}

Once we have a circuit to add a classical value to a quantum register, we can use it to build a circuit that implements the addition of a classical value modulo $p$. This adder for GF($p$) (Figure~\ref{ccaddmod}) will need to be controlled by two qubits in order to be used in a controlled multiplication circuit. After the addition of $a$ and the subtraction of $p$, we access the most significant qubit of the register. If that qubit is $\ket{1}$, then an overflow occured and we have to add back $p$ to the register. The rest of the circuit is needed to restore the ancillary qubits back to the value $\ket{0}$. Only the ADD($a$) gates need to be controlled by the control qubits because the rest of the circuit does not change the input if it is less than $p$. Since this circuit implements addition in GF($p$), we require the quantum input to be a superposition of elements of GF($p$), or integers less than $p$.

The adder for GF($p$) requires $2n+1$ qubits when not controlled and $2n+3$ when controlled by two qubits. The number of elementary gates needed is O($n$) in linear depth.

\begin{figure}[htb]
\begin{center}
\includegraphics{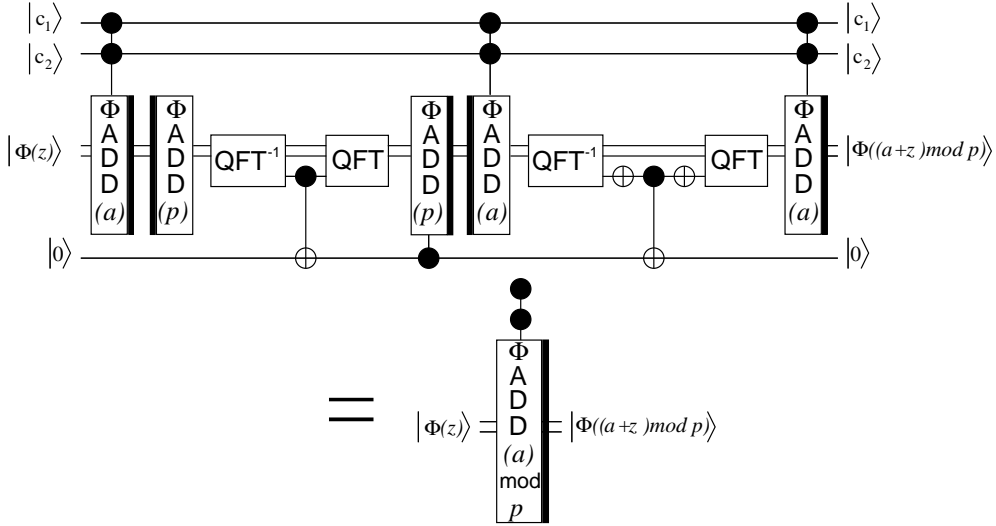}
\caption{\small{The $\phi$-adder for GF($p$) controlled by qubits $c_1$ and $c_2$. The value $a$ and the input $\ket{z}$ are both smaller than $p$.}}\label{ccphiaddmod}
\end{center}
\end{figure}

We can also build an adder for GF($p$) gate from $\phi$-adders using essentially the same method. This will be called the $\phi$-adder for GF($p$) Figure~\ref{ccphiaddmod}). It will however be necessary to use QFTs to access the most significant qubit when we need to check for the overflow and to restore that qubit. For the same reason than before, those QFTs do not have to be controlled by the control qubits, so it is possible to implement them in linear depth without ancillary
qubits.

The $\phi$-adder for GF($p$) requires $n+2$ qubits when not controlled and $n+4$ when controlled by two qubits. The number of elementary gates needed is O($n^2$) in depth O($n$).

\subsection{The controlled multiplication gate for GF($p$)}

\begin{figure}[htb]
\begin{center}
\includegraphics{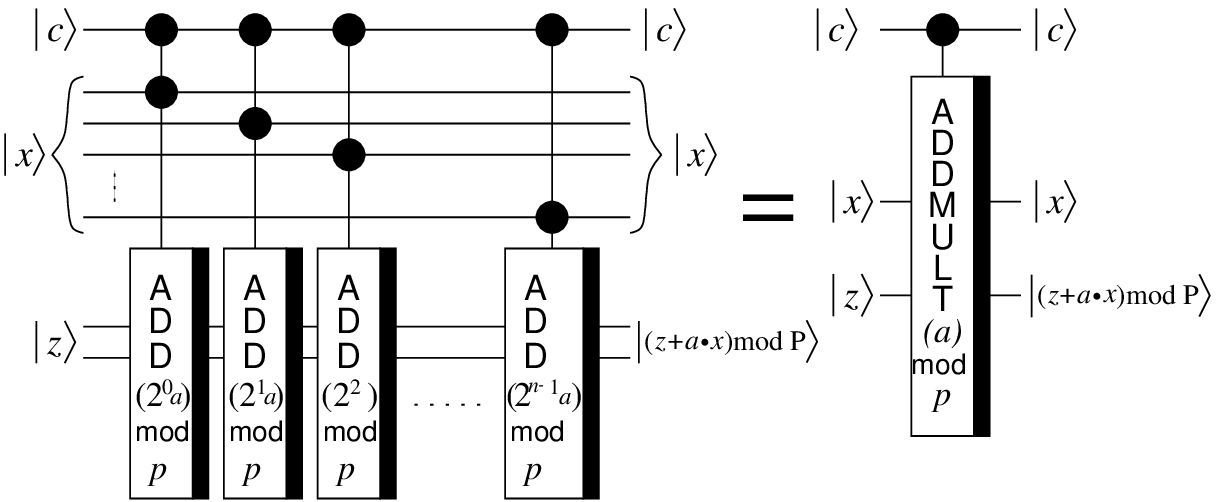}
\caption{\small{The add-mult gate. We can replace the modular adders by modular $\phi$-adders, in which case we have to apply the QFT and its inverse on the $\ket{z}$ register respectively before and after the add-mult gate.}}\label{addmult}
\end{center}
\end{figure}

Once we have a gate that adds a classical value ($a$) modulo $p$ to a quantum register $\ket{z}$, it is quite simple to implement the controlled multiplication gate. We first build a circuit that takes as input a control qubit $\ket{c}$, a quantum register $\ket{x}$ and another quantum register $\ket{z}$ which will be used as an
accumulator.  We know beforehand that the $\ket{z}$ register contains a superposition of values, all of which are smaller than $p$. Applying successive modular adders of $2^0 a$, $2^1 a$, ..., $2^n a$, the circuit leaves $\ket{c}$ and $\ket{x}$ unchanged. The $\ket{z}$ register is unchanged if $c$ = 0 and goes to $\ket{ z + a x \textrm{mod } p}$ if $c$ = 1. This will be called the add-mult gate (Fig.~\ref{addmult}).

\begin{figure}[htb]
\begin{center}
\includegraphics{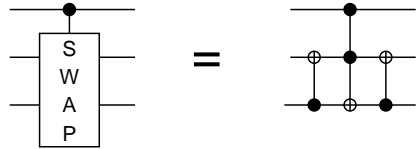}
\caption{\small{The controlled swap gate.}}\label{cswap}
\end{center}
\end{figure}

\begin{figure}[htb]
\begin{center}
\includegraphics{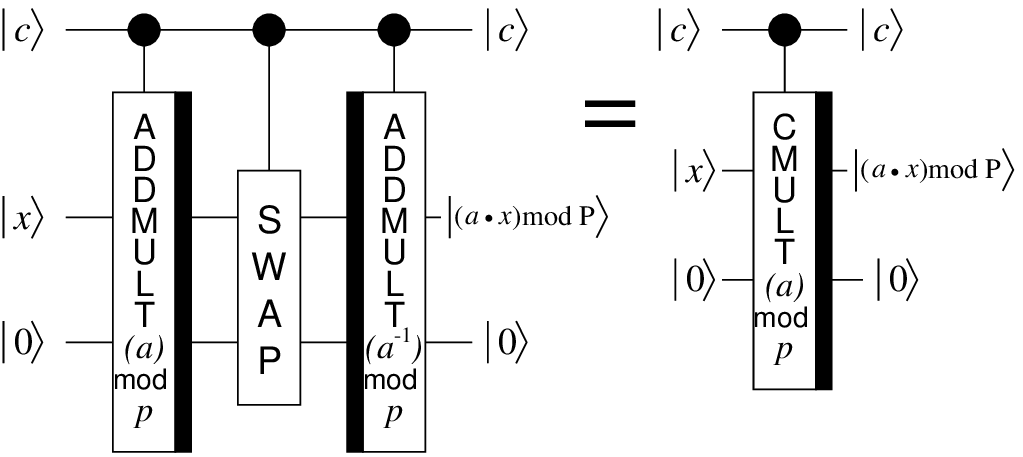}
\caption{\small{The controlled multiplication gate. The add-mult gate with a bar on the left is an inverse add-mult.}}\label{cmult}
\end{center}
\end{figure}

The problem with the add-mult circuit is that it does not take the $\ket{x}$ register directly to the product $\ket{(ax) \textrm{mod } p}$ as needed for the DLP algorithm to work. We can however use the add-mult gate to build a new circuit that does exactly what we need. We begin with a control qubit $\ket{c}$ and two registers: $\ket{x}$ and $\ket{0^n}$. After an add-mult gate, $\ket{c}$ and $\ket{x}$ are unchanged while $\ket{0^n}$ goes to $\ket{(a x)\textrm{mod } p}$ assuming $c=1$. We then apply a controlled swap gate (Figure~\ref{cswap}) to interchange $\ket{x}$ and $\ket{(a x)\textrm{mod } p}$. Finally, we apply an inverse add-mult gate with the classically computed value $a^{-1}$. The effect of this gate, always assuming $c=1$, is to leave the top register in state $\ket{(a x)\textrm{mod } p}$ while the bottom register goes to $\ket{(x- a^{-1} a x) \textrm{mod } p} = \ket{0^n} $. If $c=0$, all the registers are unchanged. We thus finally obtain the controlled multiplication gate (Figure~\ref{cmult}).

The controlled swap of two $n$-qubit registers controlled by an additional qubit thus requires $2n+1$ qubits and O($n$) elementary gates in a depth of O($n$). The swapped registers have only $n$ qubits because the extra qubit that was included to prevent overflows during the modular adder is always restored to $\ket{0}$, so it does not need to be swapped.

If the carry-sum adder is used to build the controlled multiplication gate, we need a total of $3n+2$ qubits and O($n^2$) elementary gates in a depth of O($n^2$). If the $\phi$-adder is used however, we need only $2n+3$ qubits but O($n^3$) elementary gates, again in a depth of O($n^2$).

\section{Quantum arithmetic in GF($2^n$)}

We will now consider the DLP in the Galois field GF($2^n$), so that $n$ will be the size of the inputs. The elements of GF($2^n$) can be represented as polynomials of degree $n-1$ over GF(2), or lists of $n$ bits representing the coefficients of the polynomials. The product of two elements of GF($2^n$) is the product of their polynomials modulo an irreducible fixed polynomial $Q$ of degree $n$ on GF(2). The
structure of GF($2^n$) with multiplication modulo $Q$ is independent of the choice of $Q$. Addition in GF(2) is simply the XOR operation since there is no carries in polynomial addition.

A simple way to obtain the product of two elements of GF($2^n$) is the following (see Figure~\ref{calcul}):

\begin{enumerate}
\item Given $a = (a_{n-1}, ... , a_1, a_0)$ and $x = (x_{n-1}, ... , x_1, x_0)$ where $a_i$ and $x_i$ are bits, precompute the polynomials $A_{(0)} = a ,~ A_{(1)} = (a_{n-1}, ... , a_0, 0) \textrm{mod }Q,~ A_{(2)} = (a_{n-1}, ... , a_0,0, 0) \textrm{mod }Q,~ ...~,~ A_{(n)} = (a_{n-1}, ... , a_0,\underbrace{0,...,0}_{n}) \textrm{mod }Q$. \mbox{All} those polynomials are of degree $n-1$ at most.

\item Take the product modulo 2 of $x_i$ with the coefficients of $A_{(i)}$ for $0\leq i <n$.

\item Add modulo 2 the coefficients of each polynomial $x_i A_{(i)}$ to obtain the resulting polynomial $r$.
\end{enumerate}

\begin{figure}[htb]
\begin{center}
\begin{displaymath}
\begin{array}{rrrrr}
   (a_{n-1}  & \ldots  & a_1  & a_0) & \\
   \bullet (x_{n-1}  & \ldots  & x_1  & x_0) & \\
\hline
  + x_0 \cdot [(a_{n-1}  & \ldots  & a_1  & a_0) & \textrm{mod } Q] \\
  + x_1 \cdot [(a_{n-1}~\ldots  & a_1  & a_0 & 0)& \textrm{mod } Q] \\
 +x_2 \cdot [(a_{n-1}~\ldots~a_1  & a_0 & 0 & 0)& \textrm{mod } Q] \\
  & &  \vdots & &  \\
 +x_{n-1} \cdot [(a_{n-1}~\ldots~a_1~~a_0 & 0 & \ldots & 0)& \textrm{mod } Q] \\
\hline
  (r_{n-1}  & \ldots  & r_1  & r_0) &
\end{array}
\end{displaymath}
\caption{\small{The product of two polynomials in GF($2^n$)}}\label{calcul}
\end{center}
\end{figure}

Since addition modulo 2 is simply the XOR gate, it is very easy to implement on a quantum computer. Better yet, the polynomials $a$ and $Q$ are classical values given beforehand, so we can easily precompute all the $A_{(i)}$ classically.

\subsection{The adder gate for GF($2^n$)}

\begin{figure}[htb]
\begin{center}
\includegraphics{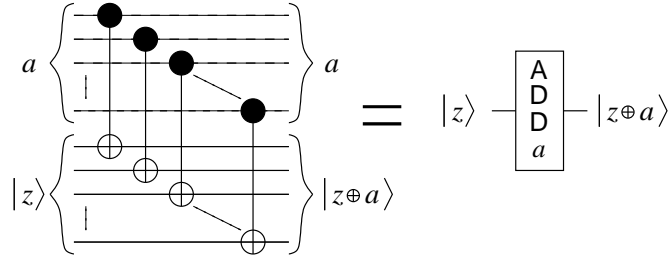}
\caption{\small{The adder for GF($2^n$). The value $a$ is classical, while $\ket{x}$ is a quantum register }}\label{adder2}
\end{center}
\end{figure}

The adder gate for GF($2^n$) is very simple. We have a quantum register $\ket{x}$ and want to add a classical value $a$ to it, so we only have to apply a NOT gate on every qubits of $\ket{x}$ corresponding to a non-zero bit of $a$ (Figure~\ref{adder2}). No gate is applied on qubits corresponding to bits of $a$ with value $0$.

This adder can be used to add any given polynomial of degree $n-1$ over GF(2), an element of GF($2^n$), to a quantum register $\ket{x}$. In general, many elements of GF($2^n$) will be in quantum superposition in this register.

The adder for GF($2^n$) is implementable directly on the $n$ qubits of the quantum register. Since there are no carries, there is no need to worry about overflows. The number of elementary gates needed is O($n$) and the depth is constant if the adder for GF($2^n$) is not controlled but changes to linear depth if it is controlled.

\subsection{The controlled multiplication gate for GF($2^n$)}

\begin{figure}[htb]
\begin{center}
\includegraphics{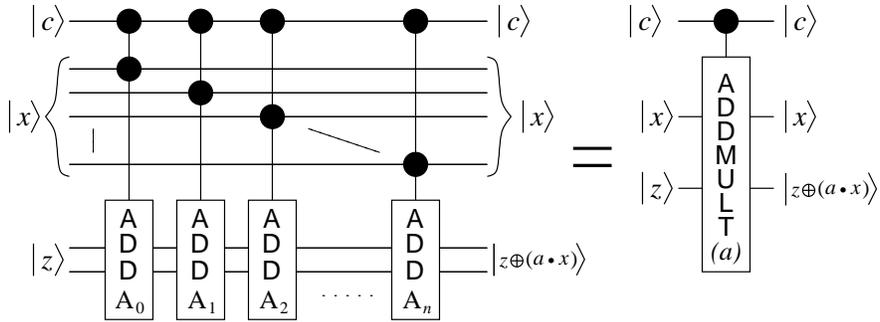}
\caption{\small{The add-mult gate for GF($2^n$).}}\label{addmult2}
\end{center}
\end{figure}

Now that we have an adder for GF($2^n$), we can carry on and build a multiplier gate for GF($2^n$). We will use essentially the same idea as with GF($p$). We build the add-mult gate by successively adding the precomputed values $A_{(0)}$, $A_{(1)}$, ... , $A_{(n)}$ to a register $\ket{z}$. The adder gate which adds the polynomial $A_{(i)}$ is controlled by the qubit $x_i$ (Figure~\ref{addmult2}).

\begin{figure}[htb]
\begin{center}
\includegraphics{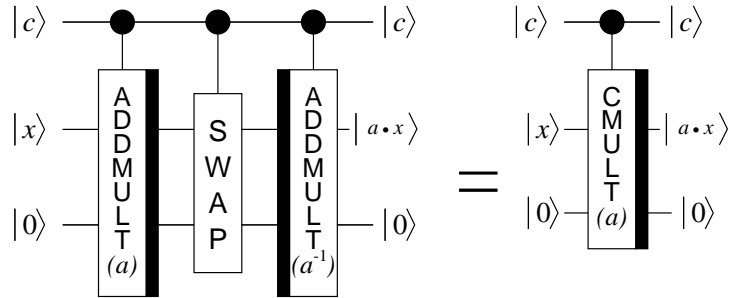}
\caption{\small{The controlled multiplication gate for GF($2^n$) and GF($p^k$).}}\label{cmult2}
\end{center}
\end{figure}

We are now set to use the trick with the controlled-swap to obtain the controlled multiplication. The idea is the same as with GF($p$).  We obtain a gate that takes a control qubit $\ket{c}$ and two registers $\ket{x}$ and $\ket{0^n}$ as inputs and outputs $\ket{c}$, $\ket{x}$ and $\ket{0^n}$ or $\ket{a \cdot x}$ depending on $c$
(Figure~\ref{cmult2}). Again, this is the only non-trivial gate needed to solve the DLP in GF($2^n$).

This controlled multiplication for GF($2^n$) requires $2n+1$ qubits and O($n^2$) gates in a depth of O($n^2$).

\section{Quantum Arithmetic in GF($p^k$)}
We now concentrate on quantum arithmetic in the Galois field GF($p^k$). For future comparison purposes, we define $n = k \ceil{\lg(p)}$, which is the size of an element in GF($p^k$). The elements of GF($p^k$) can be represented by polynomials of degree at most $k-1$ over GF($p$). They can thus be represented by lists of $k$ integers, each of these of size $\ceil{\lg(p)}$ bits for a total of $n$ bits. The product of two elements of GF($p^k$) is the product of their polynomials modulo an fixed irreducible polynomial $Q$ of degree $k$ over GF($p$). As was the case with GF($2^n$), the structure of GF($p^k$) is independent of the choice of $Q$. The product of the polynomials before reduction modulo $Q$ are taken on GF($p$), which means the coefficients are multiplied modulo $p$.

We can obtain the product of two polynomials of GF($p^k$) with the following method (see Figure~\ref{calcul2}):\label{sectionA}

\begin{enumerate}
\item Given $a = (a_{k-1}, ... , a_1, a_0)$ and $x = (x_{k-1}, ... , x_1, x_0)$ where $a_i$ and $x_i$ are numbers with $\ceil{\lg(p)}$ bits, precompute the polynomials $A_{(0)} = a,~ A_{(1)} = (a_{k-1}, ... , a_0, 0) \textrm{mod }Q,~ A_{(2)} = (a_{k-1}, ... , a_0,0, 0) \textrm{mod }Q,~...~,~A_{(k)} = (a_{k-1}, ... , a_0,\underbrace{0,...,0}_{k}) \textrm{mod }Q$. \mbox{All} those polynomials are of degree $k-1$ at most. \label{itemA}

\item Take the product modulo $p$ of $x_i$ with the coefficients of $A_{(i)}$ for $0\leq i <k$.

\item Add modulo $p$ the coefficients of each polynomial $x_i A_{(i)}$ to obtain the resulting polynomial $r$.
\end{enumerate}

\begin{figure}[htb]
\begin{center}
\begin{displaymath}
\begin{array}{rrrrll}
 (a_{k-1}  & \ldots  & a_1  & a_0) & & \\
 \bullet (x_{k-1}  & \ldots  & x_1  & x_0) & & \\
\hline
 +  [ x_0 \cdot [(a_{k-1}  & \ldots  & a_1  & a_0) & \textrm{mod } Q]& \textrm{mod } p] \\
 + [ x_1 \cdot [(a_{k-1}~ \ldots  & a_1  & a_0 & 0)& \textrm{mod } Q]& \textrm{mod } p] \\
+ [ x_2 \cdot [(a_{k-1}~ \ldots~ a_1  & a_0 & 0 & 0)& \textrm{mod } Q]& \textrm{mod } p] \\
 & & & \vdots & &  \\
 + [ x_{k-1}\cdot [(a_{k-1}~ \ldots~ a_1~~ a_0 & 0 & \ldots & 0)& \textrm{mod } Q]& \textrm{mod } p] \\
\hline
(r_{k-1}  & \ldots  & r_1  & r_0) & &
\end{array}
\end{displaymath}
\caption{\small{The product of two polynomials in GF($p^k$)}}\label{calcul2}
\end{center}
\end{figure}

As was the case for GF($2^n$), the polynomials $a$ and $Q$ are classical values given beforehand, as is the value $p$.

\subsection{The adder gate for GF($p^k$)}

The adder gate for GF($p^k$) is more complicated than that for GF($2^n$) since we will have to use adders modulo $p$ gates instead of XOR gates. We will thus make use of the adder modulo $p$ developed for the GF($p$) case.

\begin{figure}[htb]
\begin{center}
\includegraphics{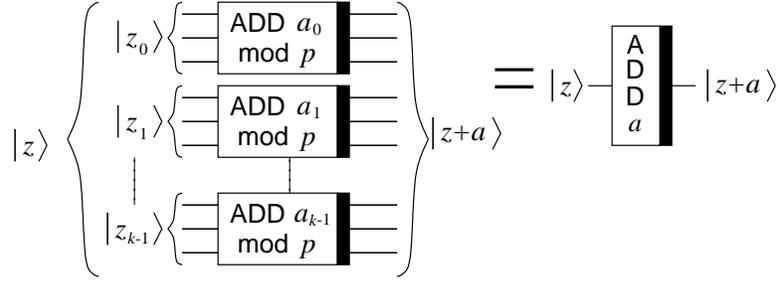}
\caption{\small{The adder for GF($p^k$). Since this gate will have to be controlled, the adder modulo $p$ gates are implemented sequentially. This also permits the recycling of ancillary qubits.}}\label{adder3}
\end{center}
\end{figure}

We want to build a gate that adds a classical value $a$ to a quantum value $\ket{z}$ where $a$ and $\ket{z}$ are elements of GF($p^k$). The quantum value $\ket{z}$ is given as an $n = k \ceil{\lg(p)}$ qubit register made of $k$ smaller registers of $\ceil{\lg(p)}$ qubits each. These sub-registers are noted $\ket{z_0}$ through $\ket{z_{k-1}}$. The
classical value $a$ is a list of $k$ numbers of $\ceil{\lg(p)}$ bits, noted $a_0$ through $a_{k-1}$. To get the adder gate for GF($p^k$), we only have to use the adder gate for GF($p$) on every sub-registers $\ket{z_i}$ to add the classical value $a_i$ to $\ket{z_i}$ (Figure~\ref{adder3}).

Of course, the input $\ket{z}$ has to be a quantum superposition of elements of GF($p^k$) for the gate to behave properly. The adders modulo $p$ used in the adder for GF($p^k$) can be made from carry-sum adders or from $\phi$-adders.

If the carry-sum adders are used as building blocks, the number of qubits needed for the GF($p^k$) adder is $n+\ceil{\lg(p)}+1$. This comes from $n=k \ceil{\lg(p)}$ qubits for the quantum input and $\ceil{\lg(p)}+1$ qubits in state $\ket{0}$ used as working space. The number of elementary gates needed is O($n$) and the depth is also O($n$).

Using the $\phi$-adders as building blocks, we need only $n+2$ qubits for the whole GF($p^k$) adder. To accomplish this however, we need to reuse two ancillary qubits for every modular $\phi$-adder gates throughout the circuit. We thus have to apply the QFT before and the invert QFT after every modular $\phi$-adder gate. This is because one
of the ancillary qubits will be used to prevent overflows, so it has to be part of the QFT. The other qubit is the one needed by the modular $\phi$-adder and is readily reusable after every modular $\phi$-adder gate. This method lets us recover the two ancillary qubits and reuse them for the next modular $\phi$-adder. The number of
elementary gates is then O($n \lg(p)$) in a depth of O($n$).

\subsection{The controlled multiplication gate for GF($p^k$)}

We now use the adder gate for GF($p^k$) to build an add-mult gate for GF($p^k$). Since the polynomials $a$ and $Q$ are classical values given beforehand, we can precalculate the values $A_{(i)}$ of item~\ref{itemA} in section~\ref{sectionA}. Furthermore, we can calculate the values $2^j A_{(i)} \textrm{mod } Q$ with $0 \leq i < k$ and $0 \leq j < \ceil{\lg(p)}$. We end up with $n$ polynomials $2^j A_{(i)} \textrm{mod } Q$, each of which is a list of $k$ integers less than $p$.

\begin{figure}[htb]
\begin{center}
\includegraphics{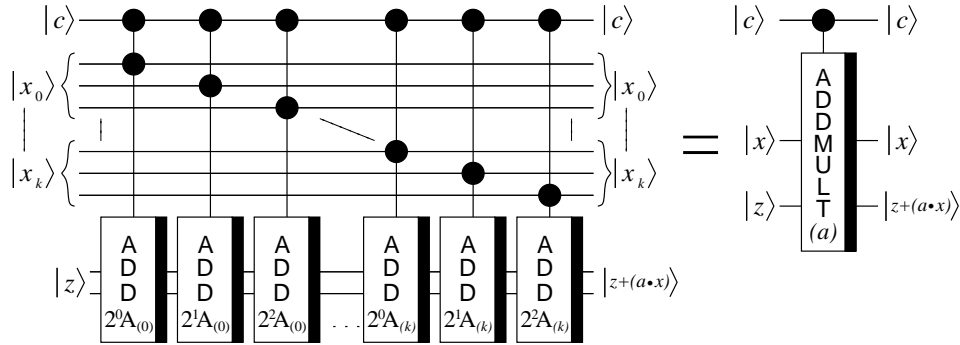}
\caption{\small{The add-mult for GF($p^k$).}}\label{addmult3}
\end{center}
\end{figure}

Each qubit of each sub-register $\ket{x_{i}}$ will control an adder gate for GF($p^k$) on the output qubits. The classical values to be added by these modular adders depend on $i$ and the position $j$ of the qubit inside $\ket{x_{i}}$. Explicitly, qubit $j$ of $\ket{x_{i}}$ will control an adder gate for GF($p^k$) on the output register where
the classical value added is the polynomial $2^j A_{(i)} \textrm{mod } Q$. The add-mult gate for GF($p^k$) consists of $k$ adder modulo $p$ gates for each qubits of $\ket{x}$ for a total of $k^2 \ceil{\lg(p)} = k n$ adder modulo $p$ gates (Figure~\ref{addmult3}).

As mentioned earlier, we are free to use either the modular adder (Figure~\ref{ccaddmod}) or the modular $\phi$-adder (Figure~\ref{ccphiaddmod}) to implement the adder modulo $p$. If we choose the latter, we have to perform the quantum Fourier transform on the output register before and after the circuit shown in figure~\ref{addmult3}.

Once we have the add-mult gate for GF($p^k$), we are in a familiar situation. We can easily use the trick with the controlled swap to get the controlled multiplication on GF($p^k$) as in figure~\ref{cmult2}.  The controlled multiplication for GF($p^k$) built from carry-sum adders requires $2n+\ceil{\lg(p)}+2$ qubits and O($n^2$) elementary gates in a depth of O($n^2$).

If $\phi$-adders are used instead, we need $2n+3$ qubits and O($n^2 \lg(p)$) elementary gates in a depth of O($n^2$) to implement the controlled multiplication for GF($p^k$).

\section{Complexity analysis}

We now compare the complexity of the controlled multiply circuits on GF($p$), GF($2^n$) and GF($p^k$). In order for the comparisons to make sense, we take $\ceil{\lg(p)} = n$ for the GF($p$) case and $k \ceil{\lg(p)} = n$ for the GF($p^k$) case.

To assess the complexity of the controlled multiplication circuits, we count the number of qubits, the number of elementary quantum gates and the depth needed for each circuit. The one qubit gates needed for these circuits are the NOT gate, the phase-shift gate and the Hadamard gate. Also needed are NOT gates controlled by up to four qubits for the circuits using the carry-sum method of addition, and phase-shifts and NOT gates controlled by one or two qubits for the circuits with the $\phi$-adders. Even though some of these gates are technologically more challenging than others, they all can be simulated by a constant number of controlled-nots and one-qubit gates~\cite{Barenco}, and are thus considered elementary. The exact gate count and depth are given in the appendix.

\begin{table}[htb]
\begin{center}
\begin{tabular}{|l||c|c|c|}
\hline
Type of adder & Width & Size & Depth\\
\hline
\hline
Carry-sum adder & {$2n$} & O($n$) & O($n$)\\
\hline
$\phi$-adder & $n+1$ &  O($n$) & 1\\
\hline
Doubly controlled carry-sum adder & {$2n+2$} & O($n$) & O($n$)\\
\hline
Doubly controlled $\phi$-adder & $n+3$ &  O($n$) & O($n$)\\
\hline
\end{tabular}
\caption{\small{The complexity of addition of inegers to quantum values without overflows.}}\label{integers}
\end{center}
\end{table}

\subsection{Controlled multiplication in GF($p$)}

For the multiplication on GF($p$), we take $p$ such that $\ceil{\lg(p)} = n$, that is $p$ is an $n$-bit prime integer.

\subsubsection{Using the carry-sum adder}

The carry-sum adder for integers that adds a classical value to a quantum one uses $2n$ qubits, O($n$) quantum gates and has a depth of O($n$). For the GF($p$) adder controlled by two qubits, we need $2n+3$ qubits, O($n$) gates and a depth of O($n$). The controlled multiplication circuit for GF($p$) thus needs $3n+2$ qubits and O($n^2$) gates in a depth of O($n^2$) with the carry-sum method of addition.

\subsubsection{Using the $\phi$-adder}

The $\phi$-adder for integers requires only $n$ qubits and O($n$) gates in constant depth if we do not count the quantum Fourier transforms.  Most of the time, the QFT are not needed before and after the $\phi$-adders because the additions are applied successively. However, we need QFTs in the $\phi$-adder for GF($p$), which takes a total of $n+4$ qubits and O($n^2$) gates in a depth of O($n$).  The controlled multiplication circuit for GF($p$) thus needs $2n+3$ qubits and O($n^3$) gates in a depth of O($n^2$) using the $\phi$-adders.

\subsection{Controlled multiplication in GF($2^n$)}

The arithmetic in GF($2^n$) are much simpler than in the other cases because we never have to worry about carries.  For GF($2^n$), the adder requires only $n$ qubits and O($n$) gates in constant depth. The doubly controlled modular adder requires $n+2$ qubits and O($n$) gates in constant depth.  The whole controlled multiplication gate requires $2n+1$ qubits and O($n^2$) gates in a depth of O($n^2$).

\subsection{Controlled multiplication in GF($p^k$)}

For the multiplication in GF($p^k$), we take $k \ceil{\lg(p)} = n$, so the elements of the field GF($p^k$) are lists of $k$ integers, each of them having at most $n/k$ bits.

\subsubsection{Using the carry-sum adder}

We don't have to build a new adder circuit for GF($p^k$) because we use $k$ adders for GF($p$). We first consider the case where these GF($p$) adders use the carry-sum method. The doubly controlled adder for GF($p^k$) requires $n+\ceil{\lg(p)}+3$ qubits and O($n$) gates in a depth of O($n$). The controlled multiplication for GF($p^k$) then requires $2n+\ceil{\lg(p)}+2$ qubits and O($n^2$) gates in a depth of O($n^2$).

\subsubsection{Using the $\phi$-adder}

We can use the $\phi$-adder for the GF($p$) additions when building the modular adders for GF($p^k$). This results in a doubly controlled adder for GF($p^k$) of $n+4$ qubits and O($n \lg(p)$) gates in a depth of O($n$). The controlled multiplication circuit then requires $2n+3$ qubits and O($n^2 \lg(p)$) = O($n^3/k$) gates in a depth of O($n^2$).

\begin{table}[htb]
\small
\begin{center}
\begin{tabular}{|p{2cm}|c|c|c|c|p{2cm}|c|}
\cline{3-7}
\multicolumn{2}{c|}{} &\multicolumn{2}{|c|}{GF($p$)} & GF($2^n$) &\multicolumn{2}{|c|}{GF($p^k$)}\\
\cline{3-4} \cline{6-7}
\multicolumn{2}{c|}{}&Carry-sum& $\phi$-adders&{}&Carry-sum& $\phi$-adders \\
\hline
Adder&Width&$2 l+1$ & $l+2$ & $n$ & $k l+k+l$ &  $k l+2$ \\
\cline{2-7}
{}&Size&O($l$) & O($l^2$)  & O($n$) & O($k l)$ &O($k l^2)$ \\
\cline{2-7}
{}&Depth&O($l$) & O($l$) & 1  & O($k l)$ & O($k l)$ \\
\hline
Doubly controlled&Width& $2l+3$ &$l+4$ & $n+2$  &$k l+k+l+2$  & $k l+4$ \\
\cline{2-7}
adder&Size&O($l$) &O($l^2$) & O($n$) & O($k l)$ &O($k l^2)$ \\
\cline{2-7}
{}&Depth&O($l$) &O($l$) & O($n$) & O($k l)$ &O($k l)$ \\
\hline
Controlled multiplication& Width & $3l+2$  & $2l+3$ & $2n+1$ & $2 k l+k+l+1$ &$2k l+3$\\
\cline{2-7}
&Size& O($l^2$) & O($l^3$) & O($n^2$) & O($k^2 l^2)$ &O($k^2 l^3)$ \\
\cline{2-7}
{}&Depth& O($l^2$) & O($l^2$) & O($n^2$) &O($k^2 l^2)$ &O($k^2 l^2)$ \\
\hline

\end{tabular}
\caption{\small{The complexity of quantum arithmetic, with $l = \ceil{\lg(p)}$.}}\label{arithmetic}
\end{center}
\end{table}
\normalsize

\section{Conclusions}
The complexity results of this paper are summarized in Table~2.  From this table, we can observe that the case GF($q$) and the GF($p^k$) for equivalent key sizes (i.e. $n=\lg q=k\lg p$) case are equivalent in terms of the quantum resources needed to implement a controlled multiplication circuit; the number of qubits required is exactly the same, and only a small constant separate the total circuit size.  This equivalence is independent of whether we choose the carry-sum adders, which minimize the number of gates, or the $\phi$-adders, which minimize the total number of qubits required.

From this, we can deduce that from the point of view of protection against quantum cryptanalytic attacks based on Shor's algorithm, no significant cryptographic advantage can be extracted from using the more complicated GF($p^k$) instead of GF($q$).  On the other hand, fewer qubits and less gates are required for the GF($2^n$) case, due to the fact that we need not keep track of carries.  Thus, there would some disadvantage in using this kind of field, in terms of protection
against quantum attacks.

\newpage
\appendix

\section{Exact complexity analysis}

We give the exact analysis of the number of elementary gates and depth of each circuit here.

\subsection{Notation}

The circuits we developed require three kinds of one-qubit gates: the NOT gate, the phase-shift gate and the Hadamard gate. These are noted respectively $\mathbf{N}$, $\mathbf{P}$ and $\mathbf{H}$. Note that for each phase-shift gates, there is a parameter by which the phase of $\ket{11}$ is multiplied and will not be explicitely taken into account in our analysis. For the circuits where the $\phi$-adders are used, we also need singly and doubly controlled $\mathbf{P}$ gates, respectively noted $\mathbf{CP}$ and $\mathbf{C^2 P}$, as well as controlled-NOT and controlled-controlled-NOT gates (or Toffoli gates), noted $\mathbf{CN}$ and $\mathbf{C^2 N}$. For the circuits where the carry-sum adders are used, NOT gates with up to four control bits are used. The notation for these gates will obviously be $\mathbf{C^3 N}$ and $\mathbf{C^4 N}$. All these gates can be simulated by a constant number of one qubit gates and $\mathbf{CN}$ gates.

\subsection{Circuits for GF($p$)}
The circuits for GF($p$) are analyzed with $n = \ceil{\lg(p)}$. We consider two different ways to implement addition on GF($p$), that is the carry-sum method and the $\phi$-adder method, and each leads to different complexity issues.

\subsubsection{The carry-sum adder for integers}
The carry-sum adder is given in figure~\ref{carrysum}. Each classical bit has a probability $\frac{1}{2}$ of being 0 and $\frac{1}{2}$ of being 1.

\begin{description}

\item[Carry (on average)] :

\begin{description}
   \item[Number of qubits =] 3

   \item[Number of gates =] $1 ~\mathbf{C^2 N} + \frac{1}{2} ~\mathbf{CN} + \frac{1}{2} ~\mathbf{N}$

   \item[Depth =] 2
 \end{description}

\item[Sum (on average)] :

\begin{description}
   \item[Number of qubits =] 2

   \item[Number of gates =] $1 ~\mathbf{CN} + \frac{1}{2} ~\mathbf{N}$

   \item[Depth =]  $\frac{3}{2}$
 \end{description}
\end{description}

The gates shown in figure~\ref{carrysum} are not all needed. The first qubit in the state $\ket{0}$ can be removed from actual implementation since it only acts as a control qubit, it will thus not be accounted for. All gates controlled by this qubit are also removed. Furthermore, we can remove some more gates that cancel each other: the two classically controlled $~\mathbf{N}$ from the top carry and inverse carry gates as well as the lone classically controlled $~\mathbf{N}$ at the bottom of the circuit with the one inside the bottommost sum gate. We are left with $(2n-3)$ carry gates, $(n-2)$ Sum gates, 2 $~\mathbf{CN}$ and $\frac{1}{2} ~\mathbf{N}$. The numbers given here are only valid for $n \geq 2$ because of the optimization of the carry-sum adder.

\begin{description}
\item[Carry-sum adder (on average)] :
\begin{description}
   \item[Number of qubits =] $2n$

   \item[Number of gates =] $(2n-3) ~\mathbf{C^2 N} + (2n-\frac{3}{2}) ~\mathbf{CN} + (\frac{3}{2}n-2) ~\mathbf{N}$

   \item[Depth =]  $\frac{11}{2}n -\frac{13}{2}$
 \end{description}
\end{description}

The singly controlled carry-sum adder is a carry-sum adder with one more qubit as a control qubits. The important thing to realize is that we only need to control the sum gates and the bottommost carry gate to get the singly controlled carry-sum adder since the other gates implement the identity if the afore-mentioned gates are removed. We end up with $(2n-4)$ carry gates, 1 controlled carry gate, $(n-2)$ controlled sum gates, 1 $\mathbf{C^2N}$ and $\frac{3}{2}~\mathbf{CN}$.

\begin{description}
\item[Singly controlled carry-sum adder (on average)] :
\begin{description}
   \item[Number of qubits =] $2n+1$

   \item[Number of gates =] $1~\mathbf{C^3 N}+(3n-\frac{9}{2})~\mathbf{C^2 N} + (\frac{3}{2}n-1) ~\mathbf{CN} + (n-2) ~\mathbf{N}$
   \item[Depth =]  $\frac{11}{2}n -\frac{13}{2}$
 \end{description}
\end{description}

The doubly controlled carry-sum adder is a carry-sum adder with two control qubits. Again, only the sum gates and the bottommost carry gate need to be controlled. We thus have $(2n-4)$ carry gates, 1 doubly controlled carry gate, $(n-2)$ doubly controlled sum gates, 1 $\mathbf{C^3N}$, $\frac{1}{2}~\mathbf{C^2N}$ and 1 $\mathbf{CN}$.

\begin{description}
\item[Doubly controlled carry-sum adder (on average)] :
\begin{description}
   \item[Number of qubits =] $2n+2$

   \item[Number of gates =] $1~\mathbf{C^4 N}+(n-\frac{1}{2})~\mathbf{C^3 N} + (\frac{5}{2}n-4) ~\mathbf{C^2 N} + (n-1) ~\mathbf{CN} + (n-2) ~\mathbf{N}$
   \item[Depth =]  $\frac{11}{2}n -\frac{13}{2}$
 \end{description}
\end{description}

\subsubsection{The adder for GF($p$)}
The adder for GF($p$) is like the doubly controlled adder of figure~\ref{ccaddmod} but without the control qubits. It consists of five carry-sum adders, one of which is controlled by a single qubit. Two $\mathbf{N}$ and two $\mathbf{CN}$ complete the circuit.

\begin{description}
\item[Adder for GF($p$)] :
\begin{description}
   \item[Number of qubits =] $2n+1$

   \item[Number of gates =] $1~\mathbf{C^3 N} + (11n-\frac{33}{2}) ~\mathbf{C^2 N} + (\frac{19}{2}n -5) ~\mathbf{CN} +(7n-8) ~\mathbf{N} $
   \item[Depth =]  $\frac{55}{2}n-\frac{57}{2}$
 \end{description}
\end{description}

\subsubsection{The doubly controlled adder for GF($p$)}
The doubly controlled adder for GF($p$) is shown in figure~\ref{ccaddmod}. It consists of five carry-sum adders, that is three which are controlled by two qubits, one which is controlled by one qubit and one which is not controlled. Two $\mathbf{N}$ and two $\mathbf{CN}$ are also needed to complete the circuit.

\begin{description}
\item[Doubly controlled adder for GF($p$)] :
\begin{description}
   \item[Number of qubits =] $2n+3$

   \item[Number of gates =] $3 ~\mathbf{C^4 N} + (3n-\frac{1}{2}) ~\mathbf{C^3 N} + (\frac{25}{2}n-\frac{39}{2}) ~\mathbf{C^2 N} + (\frac{13}{2}n -\frac{7}{2}) ~\mathbf{CN} +(\frac{11}{2}n -8) ~\mathbf{N} $

   \item[Depth =]  $\frac{55}{2}n-\frac{57}{2}$
 \end{description}
\end{description}

\subsubsection{The controlled add-mult for GF($p$)}
The controlled add-mult for GF($p$) is a modular multiplication obtained by successive modular additions (Figure~\ref{addmult}). This circuit takes as inputs a quantum value $\ket{z}$ and a number of precomputed values depending on classical value $a$. The output of the circuit is the quantum input $\ket{z}$ and a quantum register in state $\ket{z \cdot a}$ where $\cdot$ is multiplication in GF($p$). The circuit is simply $n$ doubly controlled modular adder for GF($p$) applied one after another.

\begin{description}
\item[Controlled add-mult for GF($p$)] :

\begin{description}
   \item[Number of qubits =]  $3n+2$

   \item[Number of gates =]  $3n ~\mathbf{C^4 N} + (3n^2-\frac{1}{2}n) ~\mathbf{C^3 N} + (\frac{25}{2}n^2-\frac{39}{2}n) ~\mathbf{C^2 N} + (\frac{13}{2}n^2 - \frac{7}{2}n) ~\mathbf{CN} +(\frac{11}{2}n^2-8n) ~\mathbf{N} $

   \item[Depth =] $\frac{55}{2}n^2-\frac{57}{2}n$
 \end{description}
\end{description}

\subsubsection{The controlled multiplication for GF($p$) using carry-sum adders}
The notable difference between this circuit (Figure~\ref{cmult}) and the controlled add-mult is that the controlled multiplication only outputs a register in state $\ket{z \cdot a}$ instead of keeping the input along with the output. This is important for the DLP algorithm to work properly. The controlled multiplication for GF($p$) consists of the following: a controlled add-mult of the value $a$, a swap of the $\ket{z}$ and $\ket{z \cdot a}$ registers controlled by the top qubit and finally a reverse controlled add-mult of the value $a^{-1}$. A reverse controlled add-mult is a circuit where the inverse of the gates for the controlled add-mult are applied in the reverse order.

The controlled swap of the registers are applied to swap $n$ qubits. The $~\mathbf{CN}$ can all be applied in parallel before and after the $~\mathbf{C^2 N}$ are applied.

\begin{description}
\item[Controlled swap of two registers of $n$ qubits] :
\begin{description}
   \item[Number of qubits =]  $2n+1$

   \item[Number of gates =]  $n ~\mathbf{C^2 N} + 2n \mathbf{CN}$

   \item[Depth =]$n+2$ 
 \end{description}

\item[Controlled multiplication for GF($p$) (carry-sum method)] :
\begin{description}
   \item[Number of qubits =]  $3n+2$

   \item[Number of gates =] $6n ~\mathbf{C^4 N} + (6n^2-n) ~\mathbf{C^3 N} + (25n^2-38n) ~\mathbf{C^2 N} + (13n^2-5n) ~\mathbf{CN} +(11n^2-16n)~\mathbf{N}$

   \item[Depth =]$55n^2-56n+2$ 
 \end{description}
\end{description}

\subsubsection{The $\phi$-adder for integers}
The $\phi$-adder is shown in figure~\ref{phiadd}. The $\phi$-adder takes two inputs: the quantum Fourier transform of an $n+1$ qubits register and a classical value $a$ of $n$ bits. The ($n+1$)-qubit register consists in an $n$-qubit value $\ket{z}$ with an extra leading $\ket{0}$ added in. The output will then be the quantum Fourier transform of $a+z$ in an ($n+1$)-qubit register. 

In order to compare with the results for the carry-sum adder, we consider again that each classical bits has an equal probability $\frac{1}{2}$ of being 0 or 1. Refering to figure~\ref{phiadd}, we thus have a probability $\frac{1}{2}$ of applying a gate on qubit $\ket{z_1}$, $\frac{3}{4}$ for qubit $\ket{z_2}$, $\frac{7}{8}$ for $\ket{z_3}$ and so on up to $\ket{z_n}$. The probability of applying a gate to $\ket{z_{n+1}}$ is the same as that of $\ket{z_n}$. The expected number of phase-shift gates is thus given by

\[ \sum_{k=1}^n (1-\frac{1}{2^k}) + (1-\frac{1}{2^n}) = n  \] 

for the $\phi$-adder.

\begin{description}
\item[$\phi$-adder (on average)] :
\begin{description}
   \item[Number of qubits =] $n+1$

   \item[Number of gates =]  $n ~\mathbf{P}$

   \item[Depth =] $1$ ($n$ if controlled)
 \end{description}

\item[Singly controlled $\phi$-adder (on average)] :
\begin{description}
   \item[Number of qubits =] $n+2$

   \item[Number of gates =]  $n ~\mathbf{CP}$

   \item[Depth =] $n$
 \end{description}

\item[Doubly controlled $\phi$-adder (on average)] :
\begin{description}
   \item[Number of qubits =] $n+3$

   \item[Number of gates =]  $n ~\mathbf{C^2 P}$

   \item[Depth =] $n$
 \end{description}
\end{description}

\subsubsection{The $\phi$-adder for GF($p$)}
The $\phi$-adder for GF($p$) is like the doubly controlled $\phi$-adder of figure~\ref{ccphiaddmod} but without the control qubits. In addition to the five $\phi$-adders needed for this circuit, four QFTs are also required. The sole purpose of these QFTs is to access the most significant qubit of the quantum register to detect overflows. 

\begin{description}
\item[Quantum Fourier transform on $n+1$ qubits] :

\begin{description}
   \item[Number of qubits =] $n+1$

   \item[Number of gates =] $ (\frac{n^2}{2} + \frac{n}{2})~\mathbf{CP}   + (n+1) ~\mathbf{H}$ 

   \item[Depth =] $2n+1$
 \end{description}

\item[$\phi$-adder for GF($p$)] :

\begin{description}
   \item[Number of qubits =] $n+2$

   \item[Number of gates =] $(2n^2+3n) ~\mathbf{C P} + 4n ~\mathbf{P} + 2~\mathbf{CN}+ 2~\mathbf{N}+ (4n+4) ~\mathbf{H}$

   \item[Depth =]$9n+12$
 \end{description}

\end{description}

\subsubsection{The doubly controlled $\phi$-adder for GF($p$)}
The doubly controlled modular $\phi$-adder for GF($p$) is shown in figure~\ref{ccphiaddmod}. 

\begin{description}
\item[Doubly controlled $\phi$-adder for GF($p$)] :
\begin{description}
   \item[Number of qubits =] $n+4$

   \item[Number of gates =] $3n ~\mathbf{C^2 P} + (2n^2+3n) ~\mathbf{C P} + n ~\mathbf{P} + 2~\mathbf{CN}+ 2~\mathbf{N}+ (4n+4) ~\mathbf{H}$

   \item[Depth =]$12n+9$
 \end{description}

\end{description}

\subsubsection{The controlled $\phi$-addmult for GF($p$)}
The controlled $\phi$-addmult for GF($p$) is the same as the controlled add-mult but with the carry-sum adders replaced with $\phi$-adders (figure~\ref{addmult}). A QFT and inverse QFT are needed before and after the circuit shown. These QFTs are on $n+1$ qubits.

\begin{description}
\item[Controlled $\phi$-addmult for GF($p$)] :

\begin{description}
   \item[Number of qubits =]$2n+3$

   \item[Number of gates =] $3n^2 ~\mathbf{C^2 P} + (2n^3+4n^2+n) ~\mathbf{C P} + n^2 ~\mathbf{P} + 2n~\mathbf{CN}+ 2n~\mathbf{N}+ (4n^2+6n+2) ~\mathbf{H}$

   \item[Depth =]$12n^2+13n+2$
 \end{description}
\end{description}

\subsubsection{The controlled multiplication on GF($p$) using $\phi$-adders}
The circuit is given in figure~\ref{cmult}. Since we use the $\phi$-adders this time, QFTs have to be applied before and after the $\phi$-addmult circuits.

\begin{description}
\item[Controlled-multiplication on GF($p$) with $\phi$-adders] :

\begin{description}
   \item[Number of qubits =] $2n+3$

   \item[Number of gates =] $6n^2 ~\mathbf{C^2 P} + (4n^3+8n^2+2n) ~\mathbf{C P} + 2n^2 ~\mathbf{P} + n ~\mathbf{C^2 N} + 6n~\mathbf{CN}+ 4n~\mathbf{N}+ (8n^2+12n+4) ~\mathbf{H}$

   \item[Depth =] $24n^2+27n+6$
 \end{description}
\end{description}

\subsection{Circuits for GF($2^n$)}
The circuits for GF($2^n$) are simpler than those for GF($p$) with $n = \ceil{\lg(p)}$ because there are no carries in GF($2^n$).

\subsubsection{The adder for GF($2^n$)}
The adder for GF($2^n$) is shown in figure~\ref{adder2}. The input is a quantum value $\ket{z}$ to which is added a classical value $a$, so that the output is a quantum value $\ket{z+a}$. All these values are $n$ bits or qubits long. We consider that each bit of the classical value $a$ is $1$ or $0$ with equal probabilities. 

\begin{description}
\item[Adder for GF($2^n$)] :

\begin{description}
   \item[Number of qubits =] $n$

   \item[Number of gates =]$\frac{n}{2}~\mathbf{N}$ 

   \item[Depth =]$1$
 \end{description}
\end{description}

\subsubsection{The doubly controlled adder for GF($2^n$)}
The fact that the adder has to be controlled by two qubits increases the depth and changes the gates to $\mathbf{C_2N}$.

\begin{description}
\item[Doubly controlled adder for GF($2^n$)] :

\begin{description}
   \item[Number of qubits =]$n+2$ 

   \item[Number of gates =]$\frac{n}{2}~\mathbf{C^2N}$ 

   \item[Depth =]$\frac{n}{2}$
 \end{description}
\end{description}

\subsubsection{The controlled add-mult for GF($2^n$)}
This circuit is simply a succession of $n$ adders with two control qubits (fig.~\ref{addmult2}).

\begin{description}
\item[Controlled add-mult for GF($2^n$)] :

\begin{description}
   \item[Number of qubits =] $2n+1$

   \item[Number of gates =] $\frac{n^2}{2}~\mathbf{C^2N}$ 

   \item[Depth =] $\frac{n^2}{2}$
 \end{description}
\end{description}

\subsubsection{The controlled multiplication for GF($2^n$)}
We apply the controlled add-mult of $A$, the controlled swap and the inverse controlled add-mult of $A^{-1}$ as in figure~\ref{cmult2}.

\begin{description}
\item[Controlled multiplication for GF($2^n$)] :

\begin{description}
   \item[Number of qubits =] $2n+1$ 

   \item[Number of gates =] $(n^2+n)~\mathbf{C^2N} + 2n~\mathbf{CN}$ 

   \item[Depth =] $n^2 + n + 2$
 \end{description}
\end{description}

\subsection{Circuits for GF($p^k$)}
In order to compare with the previous results, we consider GF($p^k$) with $k \ceil{\lg(p)}=n$. We don't need new adders for GF($p^k$) because we can use $k$ adders for GF($p$) in parallel instead.

\subsubsection{The adder for GF($p^k$) using carry-sum adders}
This is the adder from figure~\ref{adder3} where the modular adders use the carry-sum method. The modular adder are applied successively so that we can reuse the same ancillary qubits as work space for each modular adders. The numbers given here are valid for $p > 2$ only.

\begin{description}
\item[Adder for GF($p^k$) (carry-sum method)]:

\begin{description}
   \item[Number of qubits =] $n+k+\ceil{\lg(p)}$

   \item[Number of gates =]  $k ~\mathbf{C^3 N} + (11n-\frac{33}{2}k) ~\mathbf{C^2 N} + (\frac{19}{2}n-5k) ~\mathbf{CN} + (7n-8k)~\mathbf{N} $
   \item[Depth =] $\frac{55}{2}n-\frac{57}{2}k$
 \end{description}
\end{description}

\subsubsection{The doubly controlled adder for GF($p^k$) using carry-sum adders}
This is the previous circuit with two control qubits.

\begin{description}
\item[Doubly controlled adder for GF($p^k$) (carry-sum method)]:

\begin{description}
   \item[Number of qubits =] $n+k+\ceil{\lg(p)}+2$

   \item[Number of gates =] $3k ~\mathbf{C^4 N} + (3n-\frac{1}{2}k) ~\mathbf{C^3 N} + (\frac{25}{2}n-\frac{39}{2}k) ~\mathbf{C^2 N} + (\frac{13}{2}n-\frac{7}{2}k) ~\mathbf{CN} + (\frac{11}{2}n-8k)~\mathbf{N} $ 
   \item[Depth =] $\frac{55}{2}n-\frac{57}{2}k$
 \end{description}
\end{description}

\subsubsection{The controlled add-mult for GF($p^k$) using carry-sum adders}
This is figure~\ref{addmult3} with carry-sum adders as the building blocks. It is simply a series of $n$ instances of the previous circuit.

\begin{description}
\item[Controlled add-mult for GF($p^k$) (carry-sum method)] :

\begin{description}
   \item[Number of qubits =]  $2n+k+\ceil{\lg(p)}+1$

   \item[Number of gates =]  $3nk ~\mathbf{C^4 N} + (3n^2-\frac{1}{2}nk) ~\mathbf{C^3 N} + (\frac{25}{2}n^2-\frac{39}{2}nk) ~\mathbf{C^2 N} + (\frac{13}{2}n^2-\frac{7}{2}nk) ~\mathbf{CN} +(\frac{11}{2}n^2-8nk)~\mathbf{N} $

   \item[Depth =]  $\frac{55}{2}n^2-\frac{57}{2}nk$
 \end{description}
\end{description}

\subsubsection{The controlled multiplication for GF($p^k$) using carry-sum adders}
This is the same as what is shown in figure~\ref{cmult2} but with the add-mult for GF($p^k$).

\begin{description}
\item[Controlled multiplication for GF($p^k$) (carry-sum method)] :

\begin{description}
   \item[Number of qubits =]  $2n+k+\ceil{\lg(p)}+1$

   \item[Number of gates =] $6nk ~\mathbf{C^4 N} + (6n^2-nk) ~\mathbf{C^3 N} + (25n^2-39nk+n) ~\mathbf{C^2 N} + (13n^2 - 7nk +2n) ~\mathbf{CN}+(11n^2-16nk)~\mathbf{N} $

   \item[Depth =] $55n^2-57nk + n + 2$
 \end{description}
\end{description}

\subsubsection{The $\phi$-adder for GF($p^k$)}
This is the adder from figure~\ref{adder3} where the modular adders are $\phi$-adders. The QFT and its inverse need to be applied respectively before and after each $\phi$-adders so that we can recover the two qubits of work space needed for each modular adders. The modular adders are applied one after another.

\begin{description}
\item[$\phi$-adder for GF($p^k$)] :

\begin{description}
   \item[Number of qubits =] $n+2$

   \item[Number of gates =] $ (3n \ceil{\lg(p)}+4n) ~\mathbf{C P} + 4n ~\mathbf{P} + 2k~\mathbf{CN}+ 2k~\mathbf{N}+ (6n+6k) ~\mathbf{H}$

   \item[Depth =] $13n+14k$
 \end{description}
\end{description}

\subsubsection{The doubly controlled $\phi$-adder for GF($p^k$)}
This is the previous circuit with two control qubits.

\begin{description}
\item[Doubly controlled $\phi$-adder for GF($p^k$)] :

\begin{description}
   \item[Number of qubits =] $n+4$

   \item[Number of gates =] $3n ~\mathbf{C^2 P} + (3n \ceil{\lg(p)}+4n) ~\mathbf{C P} + n ~\mathbf{P} + 2k~\mathbf{CN}+ 2k~\mathbf{N}+ (6n+6k) ~\mathbf{H}$

   \item[Depth =] $16n+11k$
 \end{description}
\end{description}

\subsubsection{The controlled add-mult for GF($p^k$) using $\phi$-adders}
This is shown in figure~\ref{addmult3}, but this time we use $\phi$-adders as building blocks. It is again simply a series of $n$ instances of the previous circuit.

\begin{description}
\item[Controlled add-mult for GF($p^k$) with $\phi$-adders] :

\begin{description}
   \item[Number of qubits =]$2n+3$ 

   \item[Number of gates =] $3n^2 ~\mathbf{C^2 P} + (3n^2 \ceil{\lg(p)}+4n^2) ~\mathbf{C P} + n^2 ~\mathbf{P} + 2nk~\mathbf{CN}+ 2nk~\mathbf{N}+ (6n^2+6nk) ~\mathbf{H}$

   \item[Depth =] $16n^2+11nk$
 \end{description}
\end{description}

\subsubsection{The controlled multiplication for GF($p^k$) using $\phi$-adders}
Again, this is what is shown in figure~\ref{cmult2} using the $\phi$-addmult for GF($p^k$).

\begin{description}
\item[Controlled multiplication for GF($p^k$) with $\phi$-adders] :

\begin{description}
   \item[Number of qubits =]$2n+3$

   \item[Number of gates =] $6n^2 ~\mathbf{C^2 P} + (6n^2 \ceil{\lg(p)}+8n^2) ~\mathbf{C P} + 2n^2 ~\mathbf{P} + n ~\mathbf{C^2 N}+(4nk+2n)~\mathbf{CN}+ 4nk~\mathbf{N}+ (12n^2+12nk) ~\mathbf{H}$ 

   \item[Depth =] $32n^2+22nk+n+2$
 \end{description}
\end{description}

\end{document}